\pdfoutput=1
\documentclass{article}
\usepackage{spconf,amsmath,amssymb,graphicx,hyperref}
\usepackage{tikz}
\usepackage{float}
\usepackage{pifont}
\usepackage{url}
\usepackage{multirow}
\usepackage{makecell}
\usepackage{rotating}
\usepackage{booktabs}
\usepackage{xcolor}
\usepackage{xspace}
\usepackage{array}
\usepackage{caption}

\newcommand{\cmark}{\ding{51}}
\newcommand{\xmark}{\ding{55}}

\title{taming audio vaes via target-kl regularization}
\name{Prem Seetharaman* Rithesh Kumar*\thanks{*Equal contribution}}
\address{Adobe Research, San Francisco, CA}

\begin{document}
\ninept
\maketitle
\begin{abstract}
Latent diffusion models have emerged as the dominant paradigm for many generation tasks including audio generation such as text-to-audio, text-to-music and text-to-speech. A key component of latent diffusion is an autoencoder (VAE) that compresses high-dimensional signals into a low frame rate continuous representation that is conducive for downstream prediction. \textit{Regularizing} these VAEs is challenging, as there is a trade-off between over-regularized (poor output quality) and under-regularized (difficult to predict) latent representations. We propose a framework for studying this trade-off through compression and train Audio VAEs at specific bitrates via target-kl regularization. This allows direct comparison to well-studied discrete neural audio codec models, and the construction of rate-distortion curves for audio VAEs. We evaluate the impact of target-kl regularization on text-to-sound generation and find that sweeping compression rates is helpful in identifying the optimal generation setting.

\end{abstract}

\section{Introduction}

\begin{figure}[t]
\centering
\includegraphics[width=\columnwidth]{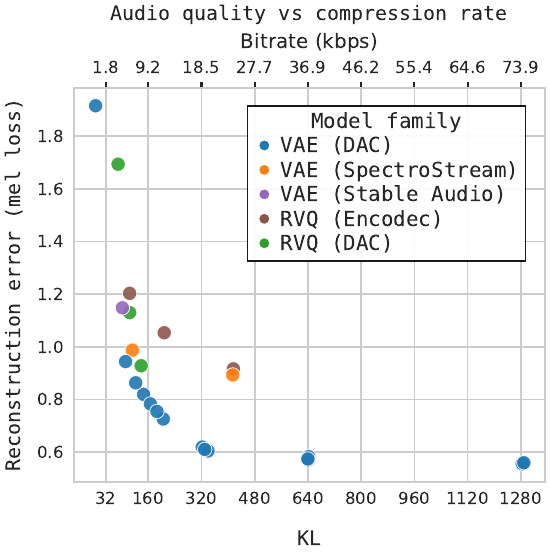}
\caption{Rate-distortion curve for different model families for both discrete and continuous audio compression. With our approach, we can target specific bitrates when training VAEs. We find that our proposed DAC-VAE achieves the best audio quality across all bitrates.}
\label{fig:kl-vs-quality}
\end{figure}

\label{sec:intro}
Hierarchical generative modeling \cite{van2017neural, rombach2022high, borsos2023audiolm, evans2025stable, wang2023neural} has become the standard approach for audio generation tasks including text-to-speech, text-to-music and text-to-sound synthesis. It involves an auto-encoder component that can compress high dimensional natural signals into low frame rate latent representations, followed by a powerful generative model that predicts the latents conditional on high-level inputs such as text. One approach uses discrete latent representations, that require training quantized variational auto-encoders (VQ-VAEs)\cite{van2017neural} followed by an autoregressive transformer token prediction model \cite{borsos2023audiolm} or their non-autoregressive counterparts \cite{garcia2023vampnet, wang2024maskgct, borsos2023soundstorm}. An alternative approach is to use a continuous latent representation that involves training gaussian-regularized VAEs \cite{kingma2013auto}, followed by a diffusion model \cite{ho2020denoising} on the latents \cite{rombach2022high, liu2023audioldm, evans2025stable}. 

The training of the autoencoder is often a dark art, with checkpoints from training runs sometimes having unpredictable behavior for generative modeling. The quality of the latent space can have an effect on the quality of any downstream generative model that uses that latent space. Well-regularized latent spaces are smooth and robust to small perturbations of the latents, which theoretically should lead to better generative models. However, over-regularized latents have poor reconstruction quality, unnecessarily putting a ceiling on the generative model. Likewise, under-regularized latents will have very good reconstruction quality, but their spaces will be sharp and sensitive to perturbations, which makes it harder for the generative model to learn the space. Often, latent spaces that are less compressed require larger generative models \cite{esser2024scaling}. We tackle these problems via:

\begin{enumerate}
  \setlength{\itemsep}{0pt}
  \setlength{\parskip}{0pt}
  \item Target-KL regularization, a novel method for targeting a specific bitrate when training a continuous VAE, which enables modelers to make trade-offs between reconstruction quality and latent regularization.
  \item A unified study of the rate-distortion trade-off for both continuous and discrete audio compression models.
  \item A study on the impact of compression rate on diffusion-based text-to-audio generative models. 
\end{enumerate}

\section{Target-kl for fixed bitrate VAE}
Autoencoders for compressing audio signals $x$ into latents $z$ are trained with the following objective:
\begin{align}
\mathbb{E}_{x\sim D}\Big[
  \mathbb{E}_{z\sim q_\phi(z|x)} \log p_\theta(x|z)
  - \lambda * D_{\mathrm{KL}}\!\big(q_\phi(z|x)\,\|\,p_\psi(z)\big)
\Big].
\end{align}

Note that when $\lambda=1$, this reduces to the original ELBO objective. In VQ-VAEs, $q_\phi(z|x)$ is deterministic and by assuming a simple uniform prior over $z$, we obtain a constant KL divergence equal to $\log K$ where $K$ is the size of the latent space. For Gaussian VAEs with approximate posterior \( q_\phi(z|x) = N(z; \mu(x), \text{diag}(\sigma^2(x))) \) and an $N(0, 1)$ prior, the KL term has a closed form: 
\begin{align}
D_{\mathrm{KL}}\!\left(q_\phi(z|x)\,\|\,p(z)\right)
&= \tfrac{1}{2}\sum_{j=1}^k \Big( \sigma_j^2 + \mu_j^2 - 1 - \log \sigma_j^2 \Big).
\end{align}

While the trade-off between compression and reconstruction quality is well studied \cite{borsos2023audiolm, zeghidour2021soundstream, kumar2023high} in (discrete) neural audio codecs — where the target bitrate is directly tied to codebook size and number of codebooks — there is no straightforward framework for reasoning about VAE spaces. 

In the context of training VAEs for latent diffusion models, a small KL penalty towards a standard normal prior $\lambda *  \mathrm{KL}\!\left( q_\phi(z \mid x) \,\|\, \mathcal{N}(0,I) \right)$ is often added as a regularizer \cite{rombach2022high} to prevent unstable or high-variance latent spaces. Selecting an appropriate weighting factor $\lambda$ for the KL term is challenging and leads to a trade-off: large values tend to over-regularize the latents,  degrading reconstruction quality, while small values yield better reconstructions but under-regularize the latents, making them challenging to predict. Current methods manually tune the KL divergence weight $\lambda$, making it difficult to systematically study this compression - reconstruction trade-off. Tuning $\lambda$ is a delicate procedure, leading to most autoencoders for latent diffusion to be regularized entirely via latent dimension size.

A key observation we make in this work is that the KL term in the ELBO also admits an interpretation in terms of coding cost or ``rate'' in the context of rate-distortion theory \cite{park2022interpreting, townsend2019practical, alemi2018information}. 
Specifically, for Gaussian VAEs the expected KL term corresponds to the average number of nats required to encode samples from the approximate posterior $q_\phi(z \mid x)$ using the prior $p(z)$ as the codebook. This viewpoint naturally links continuous latent-variable models 
to ideas from compression and quantization. The bitrate of a VAE is then (note this is a theoretical bitrate, serving as a proxy for compression):

\begin{align}
\label{eq:kl-to-kbps}
bps &= \frac{S}{\log 2} * \mathrm{KL}\!\left( q_\phi(z \mid x) \,\|\, p(z) \right)
\end{align}

where $S = \frac{f_S}{H}$ is the frame rate of the VAE, $f_S$ is the audio sample rate and $H$ is the hop length of the encoder. We can now optimize for an expected $\mathrm{KL}_{\textit{target}}$ value to obtain a VAE with a known fixed bitrate $B$:
\begin{align}
\mathrm{KL}_{\textit{target}} 
  &= \frac{B * \log 2}{S} \\
\mathcal{L}_{\textit{target-}\mathrm{KL}} 
  &= \left(\mathrm{KL} - \mathrm{KL}_{\textit{target}}\right)^2
\end{align}
 Note that in practice, we normalized both the target-KL and estimated KL by latent size $D$ since we found optimization challenging to regress the original (high) magnitude of the KL. For a bitrate $B=7.68$ kbps, latent rate of $S=40$ Hz and $D=128$, we aim for a \textit{target}-KL $\approx276$.
 
This formulation of the loss is related to the method of free bits (Eq. \ref{eq:freebits}) \cite{kingma2016improved, kingma2019introduction}, which modifies the ELBO to ensure that a certain \textit{minimum} number of bits of information are encoded per latent variable. This is achieved by clamping KL per dimension to a certain minimum value, establishing a lower bound on the overall rate budget. We find that directly regressing the KL to a target value worked better in practice.
\begin{align}
\label{eq:freebits}
\tilde{\mathcal{L}_\lambda}
&= \mathbb{E}_{x \sim D}\!\left[ \log p(x|z) \right] \notag \\
&\quad - \sum_{j=1}^K 
   \max\!\Bigl(
       \lambda,\,
       \mathbb{E}_{x \sim D}\!\left[
          D_{\mathrm{KL}}\!\bigl(q(z_j|x)\,\|\,p(z_j)\bigr)
       \right]
   \Bigr)
\end{align}

\newcommand{\xmarkgray}{\textcolor{lightgray}{\xmark}\xspace}

\begin{table}[t]
\centering
\setlength{\tabcolsep}{3pt}
\begin{tabular}{@{}p{2.0cm}|cc|cccc|ccccc@{}}
\toprule
Ablation on
& \rotatebox{90}{Passthrough} 
& \rotatebox{90}{CQT Disc} 
& \rotatebox{90}{Mel distance $\downarrow$} 
& \rotatebox{90}{KL} 
& \rotatebox{90}{Bitrate (kbps)} 
\\
\midrule

Base 
  & - & - & 0.626 & 341.34 & 19.69 \\
\midrule

\multirow{2}{*}{\makecell{Passthrough} }
  & 50\% & \xmarkgray & 0.627 & 338.56 & 19.53 \\
  & 25\% & \xmarkgray & 0.619 & 338.29 & 19.52 \\
\midrule

\multirow{2}{*}{\makecell{Discriminator}}
  & 50\% & \cmark & 0.605 & 338.72 & 19.55 \\
  & 25\% & \cmark & \textbf{0.604} & 338.61 & 19.54 \\
\bottomrule
\end{tabular}
\caption{Ablation of DAC-VAE architecture on AudioSet eval set. Passthrough refers to occasionally bypassing the bottleneck (training as a pure autoencoder). Adding a CQT discriminator and passthrough improves reconstruction quality at similar bitrates.}
\label{tab:bigtableresult}
\vspace{-1.0em}
\end{table}

\section{Experiments}

\subsection{Model architecture}
Our model is built on the same framework of neural audio codec models, except we replace the quantization bottleneck with a gaussian regularization. We use the same fully convolutional encoder-decoder model architecture from DAC \cite{kumar2023high} and the same training recipe. We train on a dataset of speech, music, and sound effects similar in composition to the original DAC recipe. In this work, we train at 48 KHz sample rate and produce a 40 Hz latent representation.

We found the original DAC recipe gives good results in the continuous setup. We make slight modifications to improve it. We add a projection of mel-spectrogram (80 mels) directly to the output of the encoder which we found to speed up model convergence. We replace the multi-band spectrogram discriminator with the CQT Discriminator proposed in BigVGAN v2 \cite{lee2022bigvgan}. Finally, training partial batches as a pure auto-encoder without any regularization \cite{defossez2024moshi} improves quality, especially for higher frequencies. We sweep two rates for this during training - 25\% and 50\%. We ablate these changes in Table \ref{tab:bigtableresult}. Note that, following Equation \ref{eq:kl-to-kbps}, we can convert the measured KL to a measured bitrate for each ablation. The actual bitrate of a model may vary slightly across ablations. In our ablations, we fix the bitrate to ~20kbps, and measure the actual bitrate when comparing models. For the rest of the paper, we fix the passthrough rate to 25\% and use the CQT discriminator for all model training. Different $\lambda$ values (1, 2, 10) result in varying degrees of adherence to the target-KL, producing the spread of points visible in Figure \ref{fig:kl-vs-quality}.

\subsection{Audio compression vs quality}
Target-KL regularization allows us to rigorously study the trade-off between reconstruction quality and latent regularization by training Audio VAEs at different fixed bitrates. We train a family of DAC-VAEs \cite{kumar2023high} with varying target-KL regularization on an internal proprietary and licensed dataset consisting of speech, music, and sound effects. Our DAC-VAE architecture is similar to the original DAC model, but with the quantization bottleneck replaced by a KL bottleneck. We evaluate all models on the evaluation subset of AudioSet \cite{gemmeke2017audio}. We set the $\mathrm{KL}_{\textit{target}}$ of our models to $80, 160, 320, 640, 1280$. We train at varying weights ($\lambda=1, 2, 10$) for the target-KL loss, which results in a variety of measured KL based on how well the model matched the target. We also train DAC-VAE with target-KL set to $0$.

In addition to DAC-VAE, we also train a continuous variant of SpectroStream \cite
{li2025spectrostream} on our own data at $\mathrm{KL}_{\textit{target}}=150, 400$.
Finally, we also evaluate existing audio autoencoder models in the literature, such as EnCodec \cite{defossez2022high}, Stable Audio VAE \cite{evans2025stable}, and DAC \cite{kumar2023high}. For DAC,
we train the model on our own data, while for the former two we rely on pretrained models. Due to our proposed conversion between bitrate and KL, we are able to compare
discrete and continuous models directly on the same plot, as seen in Figure \ref{fig:kl-vs-quality}. We train our models for 250k steps with a batch of 128 audio samples at 48khz sampling rate. For DAC models, we use $0.5$s audio segments, while for SpectroStream, we use $2$s segments.

\subsection{Text-to-audio generation}
To evaluate the downstream performance of our fixed bit-rate VAEs, we train latent diffusion models on text-to-sound and text-to-speech generation tasks. Given latent representation $z$, diffusion step $t\in(0, 1)$ and a conditioning sequence $x$ (such as text), we use v-prediction\cite{salimans2022progressive} as our model output $v_\theta(z_t, x, t)$ which predicts $v_t := \alpha_t \epsilon - \sigma_t z_t$, where $\alpha_t$ and $\sigma_t$ are defined by a noise schedule and the noised latent $z_t$ is expressed as $\alpha_t z + \sigma_t \epsilon$. We use the shifted cosine noise schedule \cite{hoogeboom2023simple, lovelace2023simple} with scale $s=log(0.5)$.

For text-to-speech, we train diffusion transformer models \cite{peebles2023scalable} to predict VAE latents conditioned on text and speaker prompt similar to DiTTo-TTS \cite{lee2024ditto, liu2024autoregressive} with a few minor differences. We use a simple decoder-only transformer architecture with 740M parameters, $d_\text{model}=1536$, $d_\text{ff}=4096$ and 12 attention heads and treat all inputs as one concatenated sequence. Instead of using a ByT5 \cite{xue2022byt5} model embeddings to represent text, we use both IPA phonemes as well as the T5 \cite{raffel2020exploring} token embeddings. In summary, our model takes T5 text embeddings, IPA phonemes, prompt latent and  noised latent as input and predicts denoised latents. For the text embedding, we use a pre-trained T5-Large model and extract phonemes using the phonemizer \cite{bernard2021phonemizer} library. All models are trained for 300k steps with AdamW optimizer, 0.01 weight decay and trained on english-only subsets of CommonVoice, Librivox and Emilia-YODAS \cite{he2025emilia} datasets. 

For text-to-sound-effect generation, we follow the same setup as text-to-speech but without the
voice cloning task. Our model generates a sound effect conditioned solely upon text, following
the same framework as \cite{kumar2024sila, garcia2025sketch2sound}. We evaluate the model on 250
hand-written text prompts and corresponding sound effects from the Adobe Audition SFX dataset \footnote{https://www.adobe.com/products/audition/offers/adobeauditiondlcsfx.html}. We evaluate audio
quality and similarity using FLAM \cite{wu2025flam}. All models are trained for 400k steps, and have
~1B parameters, on a proprietary and licensed dataset of sound effects. Our model is a 24-layer transformer, with $d_{model}=1536, d_{ff}=4096$ and 12 attention heads and SwiGLU activations. We cross-attend to text embeddings from T5-XXL \cite{chung2024scaling}. Across experiments, we only vary the VAE that is used to convert audio into latents, keeping all other hyperparameters fixed.

\begin{table}[t]
\centering
\caption{Results of diffusion transformer applied to the text-to-sound-effects task, on a variety of VAEs. We measure text-audio similarity and quality using FLAM \cite{wu2025flam}, where FAD and KAD measure distribution-level quality. Target KL values were 80, 160, 320, 640, 1280; measured KL may differ. We also show Stable Audio Open (SAO), which uses a different VAE.}
\label{tab:tta-metrics}
\begin{tabular}{@{}l|ll|lll@{}}
\toprule
    Model & KL & Bitrate & Text-audio sim. & KAD & FAD \\
    \midrule
    Ours & 132.63 & 7.65 & 69.76 & 1.93 & 0.11 \\
    Ours & 200.39 & 11.56 & \textbf{70.67} & \textbf{1.70} & \textbf{0.11} \\
    Ours & 341.26 & 19.69 & 68.80 & 2.28 & 0.12 \\
    Ours & 642.35 & 37.06 & 68.99 & 2.02 & 0.12 \\ 
    Ours & 1284.21 & 74.10 & 66.84 & 2.16 & 0.12 \\
    \midrule
    SAO & 82.16 & 4.74 & 68.38 & 2.13 & 0.13 \\
  \bottomrule
\end{tabular}
\end{table}

\begin{table}[t]
\centering
\caption{Results of our diffusion model trained for TTS and evaluated on the SEED-en \cite{anastassiou2024seed} test sets. WER is measured using Whisper Large-v3 transcription and SSIM is measured using WavLM-based speaker embedding model.}
\label{tab:tts-metrics}
\begin{tabular}{@{}ll|ll@{}}
\toprule
    KL & Bitrate (kbps) & WER & SSIM\\
    \midrule
132.63 & 7.65 & \textbf{1.61} & \textbf{0.68}\\
200.39 & 11.56 & 1.70 & 0.68\\
341.26 & 19.69 & 1.98 & 0.67\\
642.35 & 37.06 & 1.75 & 0.66\\ 
1284.21 & 74.10 & 1.61 & 0.67\\
\bottomrule
\end{tabular}
\end{table}

\section{Results}

In Figure \ref{fig:kl-vs-quality}, we show the rate-distortion trends for 
a variety of discrete and continuous audio compression models. We find that
target-KL regularization allows us to target specific bitrates for continuous
VAEs and study how various architectures behave under different compression 
rates explicitly. We find that DAC-VAE seems to form the pareto curve for
audio compression, achieving the best reconstruction quality across all bitrates. In addition, we can do a direct comparison to well-studied discrete models such as EnCodec and DAC. We find that while the DAC implementation
is almost identical to DAC-VAE, the discrete quantization bottleneck does not scale with bitrate as effectively as a continuous bottleneck (note that DAC-RVQ uses structured dropout to support multiple bitrates from a single model, while we train separate VAE models per target-KL). This drawback of vector-quantization was also noticed in prior work \cite{mentzer2023finite}.

We find that other VAEs - SpectroStream and Stable Audio VAE - do not perform as well as DAC-VAE at similar bitrates, and both are ``off'' the pareto curve. By comparing VAEs at specific bitrates, we can make more principled comparisons between architectures and identify the best model family for a given bitrate. While SpectroStream comes close to the performance of DAC-VAE in the low-bitrate regime, it does not scale as well to higher bitrates. Our goal here is to demonstrate how target-KL regularization allows for principled and systematic comparison of different architectures for audio VAEs, as done in discrete codec literature \cite{defossez2022high, kumar2023high}.

In Table \ref{tab:tta-metrics}, we show the results of our text-to-sound-effect diffusion model trained on different VAEs. We find that the model trained with the $~200$ target-KL ($11.56$ kbps) DAC-VAE achieves the best performance across all metrics, with a text-audio similarity score of 70.67 and KAD of 1.70. We find that both lower and higher bitrate models perform worse, indicating that there is an optimal compression rate for this task. We hypothesize that lower bitrate models are too over-regularized, leading to poor reconstruction quality, while higher bitrate models are under-regularized, making them difficult to predict. Sweeping the compression rate allows us to identify this optimal point, and eliminates guesswork in training the autoencoder prior to the generative modeling stage. We also find that Stable Audio Open \cite{evans2025stable}, which uses a different VAE with more compression produces metrics that follow a similar trend to our sweep, with slightly lower text-audio similarity and worse quality scores. 

We also observed an intriguing pattern when training text-to-speech diffusion models on our VAEs, as summarized in Table \ref{tab:tts-metrics}. Diffusion models trained on lower-bitrate VAEs generally achieved lower word error rates (WER) and higher speaker similarity (SSIM), whereas higher bitrates often led to increased pronunciation errors. Interestingly, we also noted an exception: certain high-bitrate VAEs produced lower WER. However, qualitative inspection of these samples revealed that  despite accurate content, the speech sounded less natural and more monotonous than those generated at lower bitrates. In contrast to the task of text-to-audio generation, diffusion models for TTS are able to directly copy more information from the prompt audio latents (especially if unregularized and carry more information) that complicate this analysis. We leave a detailed investigation of this phenomenon to future work, particularly with evaluation metrics that better capture semantic characteristics such as prosody and naturalness. However, we find that a low bitrate VAE (11.56kbps) performs well across both tasks, indicating that this compression rate is a good starting point for training audio VAEs for generative modeling.

\section{Conclusion and Future work}
In this work, we proposed target-KL regularization, a method for training continuous VAEs at fixed bitrates. This allows for direct comparison to discrete neural audio codecs and enables systematic study of the rate-distortion trade-off for continuous audio compression models. We evaluated our models on text-to-sound and text-to-speech generation tasks, finding that sweeping the compression rate is helpful in identifying the optimal generation setting. In addition, we adapted DAC to the continuous setting, and made small improvements to the training recipe. We find that our proposed VAE forms a pareto frontier for audio compression models. 

Our proposed framework provides a solid foundation for 
future work in audio autoencoders. As an example, all of the DAC models trained in this work had the same number of parameters. In this framework, we can now ask: given a fixed bitrate budget for the VAE, how can scaling model size affect reconstruction quality? Similarily, we can investigate the interaction between the rate of the latent space and the compression amount, or the latent dimension and the compression rate. Finally, our regularization can be combined with other techniques like semantic alignment \cite{defossez2024moshi, yu2024representation} for further research into how VAEs can be best trained for downstream generative tasks. 

\bibliographystyle{IEEEbib}
\bibliography{strings,refs}

@article{kingma2016improved,
  title={Improved variational inference with inverse autoregressive flow},
  author={Kingma, Durk P and Salimans, Tim and Jozefowicz, Rafal and Chen, Xi and Sutskever, Ilya and Welling, Max},
  journal={Advances in neural information processing systems},
  volume={29},
  year={2016}
}

@article{anastassiou2024seed,
  title={Seed-tts: A family of high-quality versatile speech generation models},
  author={Anastassiou, Philip and Chen, Jiawei and Chen, Jitong and Chen, Yuanzhe and Chen, Zhuo and Chen, Ziyi and Cong, Jian and Deng, Lelai and Ding, Chuang and Gao, Lu and others},
  journal={arXiv preprint arXiv:2406.02430},
  year={2024}
}

@inproceedings{wu2025flam,
title={{FLAM}: Frame-Wise Language-Audio Modeling},
author={Yusong Wu and Christos Tsirigotis and Ke Chen and Cheng-Zhi Anna Huang and Aaron Courville and Oriol Nieto and Prem Seetharaman and Justin Salamon},
booktitle={Forty-second International Conference on Machine Learning},
year={2025},
url={https://openreview.net/forum?id=7fQohcFrxG}}

@inproceedings{gemmeke2017audio,
  title={Audio set: An ontology and human-labeled dataset for audio events},
  author={Gemmeke, Jort F and Ellis, Daniel PW and Freedman, Dylan and Jansen, Aren and Lawrence, Wade and Moore, R Channing and Plakal, Manoj and Ritter, Marvin},
  booktitle={2017 IEEE international conference on acoustics, speech and signal processing (ICASSP)},
  pages={776--780},
  year={2017},
  organization={IEEE}
}

@article{li2025spectrostream,
  title={SpectroStream: A Versatile Neural Codec for General Audio},
  author={Li, Yunpeng and Han, Kehang and McWilliams, Brian and Borsos, Zalan and Tagliasacchi, Marco},
  journal={arXiv preprint arXiv:2508.05207},
  year={2025}
}

@article{kumar2024sila,
  title={SILA: Signal-to-Language Augmentation for Enhanced Control in Text-to-Audio Generation},
  author={Kumar, Sonal and Seetharaman, Prem and Salamon, Justin and Manocha, Dinesh and Nieto, Oriol},
  journal={arXiv preprint arXiv:2412.09789},
  year={2024}
}

@inproceedings{garcia2025sketch2sound,
  title={Sketch2sound: Controllable audio generation via time-varying signals and sonic imitations},
  author={Garc{\'\i}a, Hugo Flores and Nieto, Oriol and Salamon, Justin and Pardo, Bryan and Seetharaman, Prem},
  booktitle={ICASSP 2025-2025 IEEE International Conference on Acoustics, Speech and Signal Processing (ICASSP)},
  pages={1--5},
  year={2025},
  organization={IEEE}
}

@article{chung2024scaling,
  title={Scaling instruction-finetuned language models},
  author={Chung, Hyung Won and Hou, Le and Longpre, Shayne and Zoph, Barret and Tay, Yi and Fedus, William and Li, Yunxuan and Wang, Xuezhi and Dehghani, Mostafa and Brahma, Siddhartha and others},
  journal={Journal of Machine Learning Research},
  volume={25},
  number={70},
  pages={1--53},
  year={2024}
}

@article{mentzer2023finite,
  title={Finite scalar quantization: Vq-vae made simple},
  author={Mentzer, Fabian and Minnen, David and Agustsson, Eirikur and Tschannen, Michael},
  journal={arXiv preprint arXiv:2309.15505},
  year={2023}
}

@article{yu2024representation,
  title={Representation alignment for generation: Training diffusion transformers is easier than you think},
  author={Yu, Sihyun and Kwak, Sangkyung and Jang, Huiwon and Jeong, Jongheon and Huang, Jonathan and Shin, Jinwoo and Xie, Saining},
  journal={arXiv preprint arXiv:2410.06940},
  year={2024}
}

@inproceedings{esser2024scaling,
  title={Scaling rectified flow transformers for high-resolution image synthesis},
  author={Esser, Patrick and Kulal, Sumith and Blattmann, Andreas and Entezari, Rahim and M{\"u}ller, Jonas and Saini, Harry and Levi, Yam and Lorenz, Dominik and Sauer, Axel and Boesel, Frederic and others},
  booktitle={Forty-first international conference on machine learning},
  year={2024}
}

@article{kumar2023high,
  title={High-fidelity audio compression with improved rvqgan},
  author={Kumar, Rithesh and Seetharaman, Prem and Luebs, Alejandro and Kumar, Ishaan and Kumar, Kundan},
  journal={Advances in Neural Information Processing Systems},
  volume={36},
  pages={27980--27993},
  year={2023}
}

@article{defossez2022high,
  title={High fidelity neural audio compression},
  author={D{\'e}fossez, Alexandre and Copet, Jade and Synnaeve, Gabriel and Adi, Yossi},
  journal={arXiv preprint arXiv:2210.13438},
  year={2022}
}

@inproceedings{rombach2022high,
  title={High-resolution image synthesis with latent diffusion models},
  author={Rombach, Robin and Blattmann, Andreas and Lorenz, Dominik and Esser, Patrick and Ommer, Bj{\"o}rn},
  booktitle={Proceedings of the IEEE/CVF conference on computer vision and pattern recognition},
  pages={10684--10695},
  year={2022}
}

@article{park2022interpreting,
  title={Interpreting rate-distortion of variational autoencoder and using model uncertainty for anomaly detection},
  author={Park, Seonho and Adosoglou, George and Pardalos, Panos M},
  journal={Annals of Mathematics and Artificial Intelligence},
  volume={90},
  number={7},
  pages={735--752},
  year={2022},
  publisher={Springer}
}

@article{townsend2019practical,
  title={Practical lossless compression with latent variables using bits back coding},
  author={Townsend, James and Bird, Tom and Barber, David},
  journal={arXiv preprint arXiv:1901.04866},
  year={2019}
}

@article{alemi2018information,
  title={An information-theoretic analysis of deep latent-variable models},
  author={Alemi, Alex and Poole, Ben and Fischer, Ian and Dillon, Josh and Saurus, Rif A and Murphy, Kevin},
  journal={arXiv preprint arXiv:1711.00464},
  year={2018}
}

@article{kingma2019introduction,
  title={An introduction to variational autoencoders},
  author={Kingma, Diederik P and Welling, Max and others},
  journal={Foundations and Trends{\textregistered} in Machine Learning},
  volume={12},
  number={4},
  pages={307--392},
  year={2019},
  publisher={Now Publishers, Inc.}
}

@article{lee2022bigvgan,
  title={Bigvgan: A universal neural vocoder with large-scale training},
  author={Lee, Sang-gil and Ping, Wei and Ginsburg, Boris and Catanzaro, Bryan and Yoon, Sungroh},
  journal={arXiv preprint arXiv:2206.04658},
  year={2022}
}

@article{defossez2024moshi,
  title={Moshi: a speech-text foundation model for real-time dialogue},
  author={D{\'e}fossez, Alexandre and Mazar{\'e}, Laurent and Orsini, Manu and Royer, Am{\'e}lie and P{\'e}rez, Patrick and J{\'e}gou, Herv{\'e} and Grave, Edouard and Zeghidour, Neil},
  journal={arXiv preprint arXiv:2410.00037},
  year={2024}
}

@article{lee2024ditto,
  title={Ditto-tts: Efficient and scalable zero-shot text-to-speech with diffusion transformer},
  author={Lee, Keon and Kim, Dong Won and Kim, Jaehyeon and Cho, Jaewoong},
  journal={arXiv e-prints},
  pages={arXiv--2406},
  year={2024}
}

@article{liu2024autoregressive,
  title={Autoregressive diffusion transformer for text-to-speech synthesis},
  author={Liu, Zhijun and Wang, Shuai and Inoue, Sho and Bai, Qibing and Li, Haizhou},
  journal={arXiv preprint arXiv:2406.05551},
  year={2024}
}

@inproceedings{liu2023audioldm,
  title={{AudioLDM}: Text-to-audio generation with latent diffusion models},
  author={Liu, Haohe and Chen, Zehua and Yuan, Yi and Mei, Xinhao and Liu, Xubo and Mandic, Danilo and Wang, Wenwu and Plumbley, Mark D},
  booktitle={Proceedings of the 40th International Conference on Machine Learning (ICML)},
  year={2023}
}

@inproceedings{peebles2023scalable,
  title={Scalable diffusion models with transformers},
  author={Peebles, William and Xie, Saining},
  booktitle={Proceedings of the IEEE/CVF international conference on computer vision},
  pages={4195--4205},
  year={2023}
}

@article{xue2022byt5,
  title={ByT5: Towards a token-free future with pre-trained byte-to-byte models},
  author={Xue, Linting and Barua, Aditya and Constant, Noah and Al-Rfou, Rami and Narang, Sharan and Kale, Mihir and Roberts, Adam and Raffel, Colin},
  journal={Transactions of the Association for Computational Linguistics},
  volume={10},
  pages={291--306},
  year={2022},
  publisher={MIT Press One Broadway, 12th Floor, Cambridge, Massachusetts 02142, USA~…}
}

@article{raffel2020exploring,
  title={Exploring the limits of transfer learning with a unified text-to-text transformer},
  author={Raffel, Colin and Shazeer, Noam and Roberts, Adam and Lee, Katherine and Narang, Sharan and Matena, Michael and Zhou, Yanqi and Li, Wei and Liu, Peter J},
  journal={Journal of machine learning research},
  volume={21},
  number={140},
  pages={1--67},
  year={2020}
}

@article{van2017neural,
  title={Neural discrete representation learning},
  author={Van Den Oord, Aaron and Vinyals, Oriol and others},
  journal={Advances in neural information processing systems},
  volume={30},
  year={2017}
}

@article{borsos2023audiolm,
  title={Audiolm: a language modeling approach to audio generation},
  author={Borsos, Zal{\'a}n and Marinier, Rapha{\"e}l and Vincent, Damien and Kharitonov, Eugene and Pietquin, Olivier and Sharifi, Matt and Roblek, Dominik and Teboul, Olivier and Grangier, David and Tagliasacchi, Marco and others},
  journal={IEEE/ACM transactions on audio, speech, and language processing},
  volume={31},
  pages={2523--2533},
  year={2023},
  publisher={IEEE}
}

@inproceedings{evans2025stable,
  title={Stable audio open},
  author={Evans, Zach and Parker, Julian D and Carr, CJ and Zukowski, Zack and Taylor, Josiah and Pons, Jordi},
  booktitle={ICASSP 2025-2025 IEEE International Conference on Acoustics, Speech and Signal Processing (ICASSP)},
  pages={1--5},
  year={2025},
  organization={IEEE}
}

@article{wang2023neural,
  title={Neural codec language models are zero-shot text to speech synthesizers},
  author={Wang, Chengyi and Chen, Sanyuan and Wu, Yu and Zhang, Ziqiang and Zhou, Long and Liu, Shujie and Chen, Zhuo and Liu, Yanqing and Wang, Huaming and Li, Jinyu and others},
  journal={arXiv preprint arXiv:2301.02111},
  year={2023}
}

@article{wang2024maskgct,
  title={Maskgct: Zero-shot text-to-speech with masked generative codec transformer},
  author={Wang, Yuancheng and Zhan, Haoyue and Liu, Liwei and Zeng, Ruihong and Guo, Haotian and Zheng, Jiachen and Zhang, Qiang and Zhang, Xueyao and Zhang, Shunsi and Wu, Zhizheng},
  journal={arXiv preprint arXiv:2409.00750},
  year={2024}
}

@article{garcia2023vampnet,
  title={Vampnet: Music generation via masked acoustic token modeling},
  author={Garcia, Hugo Flores and Seetharaman, Prem and Kumar, Rithesh and Pardo, Bryan},
  journal={arXiv preprint arXiv:2307.04686},
  year={2023}
}

@article{borsos2023soundstorm,
  title={Soundstorm: Efficient parallel audio generation},
  author={Borsos, Zal{\'a}n and Sharifi, Matt and Vincent, Damien and Kharitonov, Eugene and Zeghidour, Neil and Tagliasacchi, Marco},
  journal={arXiv preprint arXiv:2305.09636},
  year={2023}
}

@article{kingma2013auto,
  title={Auto-encoding variational bayes},
  author={Kingma, Diederik P and Welling, Max},
  journal={arXiv preprint arXiv:1312.6114},
  year={2013}
}

@article{ho2020denoising,
  title={Denoising diffusion probabilistic models},
  author={Ho, Jonathan and Jain, Ajay and Abbeel, Pieter},
  journal={Advances in neural information processing systems},
  volume={33},
  pages={6840--6851},
  year={2020}
}

@article{zeghidour2021soundstream,
  title={Soundstream: An end-to-end neural audio codec},
  author={Zeghidour, Neil and Luebs, Alejandro and Omran, Ahmed and Skoglund, Jan and Tagliasacchi, Marco},
  journal={IEEE/ACM Transactions on Audio, Speech, and Language Processing},
  volume={30},
  pages={495--507},
  year={2021},
  publisher={IEEE}
}

@article{salimans2022progressive,
  title={Progressive distillation for fast sampling of diffusion models},
  author={Salimans, Tim and Ho, Jonathan},
  journal={arXiv preprint arXiv:2202.00512},
  year={2022}
}

@article{lovelace2023simple,
  title={Simple-TTS: End-to-end text-to-speech synthesis with latent diffusion},
  author={Lovelace, Justin and Ray, Soham and Kim, Kwangyoun and Weinberger, Kilian Q and Wu, Felix},
  journal={arXiv preprint},
  year={2023}
}

@article{bernard2021phonemizer,
  title={Phonemizer: Text to phones transcription for multiple languages in python},
  author={Bernard, Mathieu and Titeux, Hadrien},
  journal={Journal of Open Source Software},
  volume={6},
  number={68},
  pages={3958},
  year={2021}
}

@inproceedings{hoogeboom2023simple,
  title={simple diffusion: End-to-end diffusion for high resolution images},
  author={Hoogeboom, Emiel and Heek, Jonathan and Salimans, Tim},
  booktitle={International Conference on Machine Learning},
  pages={13213--13232},
  year={2023},
  organization={PMLR}
}

@article{he2025emilia,
  title={Emilia: A large-scale, extensive, multilingual, and diverse dataset for speech generation},
  author={He, Haorui and Shang, Zengqiang and Wang, Chaoren and Li, Xuyuan and Gu, Yicheng and Hua, Hua and Liu, Liwei and Yang, Chen and Li, Jiaqi and Shi, Peiyang and others},
  journal={arXiv preprint arXiv:2501.15907},
  year={2025}
}

\end{document}